\documentclass[twocolumn,english,aps,prl, raggedbottom,showpacs,floatfix,longbibliography]{revtex4}
\usepackage[T1]{fontenc}
\usepackage[utf8]{inputenc}
\usepackage{babel}
\usepackage{verbatim}
\usepackage{amsmath}
\usepackage{amssymb}
\usepackage{graphicx}
\usepackage{esint}
\usepackage{hyperref}
\usepackage{breakurl}

\makeatletter
\@ifundefined{textcolor}{}
{%
 \definecolor{BLACK}{gray}{0}
 \definecolor{WHITE}{gray}{1}
 \definecolor{RED}{rgb}{1,0,0}
 \definecolor{GREEN}{rgb}{0,1,0}
 \definecolor{BLUE}{rgb}{0,0,1}
 \definecolor{CYAN}{cmyk}{1,0,0,0}
 \definecolor{MAGENTA}{cmyk}{0,1,0,0}
 \definecolor{YELLOW}{cmyk}{0,0,1,0}
}

\usepackage{color}
\usepackage{babel}
\usepackage{soul}

\@ifundefined{textcolor}{}{%
 \definecolor{BLACK}{gray}{0}
 \definecolor{WHITE}{gray}{1}
 \definecolor{RED}{rgb}{1,0,0}
 \definecolor{GREEN}{rgb}{0,1,0}
 \definecolor{BLUE}{rgb}{0,0,1}
 \definecolor{CYAN}{cmyk}{1,0,0,0}
 \definecolor{MAGENTA}{cmyk}{0,1,0,0}
 \definecolor{YELLOW}{cmyk}{0,0,1,0}
}

\@ifundefined{definecolor}{color}{}
\usepackage{babel}

\makeatother

\begin{document}

\title{Haldane phase in one-dimensional topological Kondo insulators}

\author{Alejandro Mezio}

\affiliation{Facultad de Ciencias Exactas, Ingenier\'ia y Agrimensura, Universidad
Nacional de Rosario and Instituto de F\'isica Rosario, Bv. 27 de
Febrero 210 bis, 2000 Rosario, Argentina}

\author{Alejandro M. Lobos}

\email{lobos@ifir-conicet.gov.ar}

\affiliation{Facultad de Ciencias Exactas, Ingenier\'ia y Agrimensura, Universidad
Nacional de Rosario and Instituto de F\'isica Rosario, Bv. 27 de
Febrero 210 bis, 2000 Rosario, Argentina}

\author{Ariel O. Dobry}

\affiliation{Facultad de Ciencias Exactas, Ingenier\'ia y Agrimensura, Universidad
Nacional de Rosario and Instituto de F\'isica Rosario, Bv. 27 de
Febrero 210 bis, 2000 Rosario, Argentina}

\author{Claudio J. Gazza}

\affiliation{Facultad de Ciencias Exactas, Ingenier\'ia y Agrimensura, Universidad
Nacional de Rosario and Instituto de F\'isica Rosario, Bv. 27 de
Febrero 210 bis, 2000 Rosario, Argentina}

\date{\today}
\begin{abstract}
We investigate the groundstate properties of a recently proposed model
for a topological Kondo insulator in one dimension (i.e., the $p$-wave
Kondo-Heisenberg lattice model) by means of the Density Matrix Renormalization
Group method. The non-standard Kondo interaction in this model is
different from the usual (i.e., local) Kondo interaction in that the
localized spins couple to the ``$p$-wave'' spin density of conduction
electrons, inducing a topologically non-trivial insulating groundstate.
Based on the analysis of the charge- and spin-excitation gaps, the
string order parameter, and the spin profile in the groundstate, we show that, at half-filling
and low energies, the system is in the Haldane phase and hosts
topologically protected spin-1/2 end-states. Beyond its intrinsic
interest as a useful ``toy-model'' to understand the effects of
strong correlations on topological insulators, we show that the $p$-wave
Kondo-Heisenberg model can be implemented in $p-$band optical lattices
loaded with ultra-cold Fermi gases. 
\end{abstract}


\maketitle
{} %

\textit{Introduction.} Topological Kondo insulators (TKI) are a type
of recently proposed materials where strong interactions and topology
naturally coexist \citep{Dzero10_Topological_Kondo_Insulators,Dzero12_Theory_of_topological_Kondo_insulators,Dzero15_Review_TKIs}. Within a mean-field picture \citep{read83,coleman87,newns87}, TKIs can
be understood as a strongly renormalized $f$-electron band lying
close to the Fermi level, and hybridizing with the conduction-electron
$d-$bands.
At half-filling, an insulating state appears due to the opening of
a low-temperature hybridization gap induced by interactions at the
Fermi energy. Due to the opposite parities of the states being hybridized,
a topologically non-trivial ground state emerges, characterized by
an insulating gap in the bulk and conducting Dirac states at the surface.
At present, TKI materials, among which samarium hexaboride (SmB$_{6}$)
is the best known example, are under intense investigation both theoretically
and experimentally \citep{Wolgast13_SS_in_SmB6,Zhang13_SS_in_SmB6,Xu14_SARPES_in_SmB6,Kim14_Topological_SS_in_SmB6} 

\begin{figure}
\includegraphics[clip,scale=0.45, bb = 20 30 739 210]{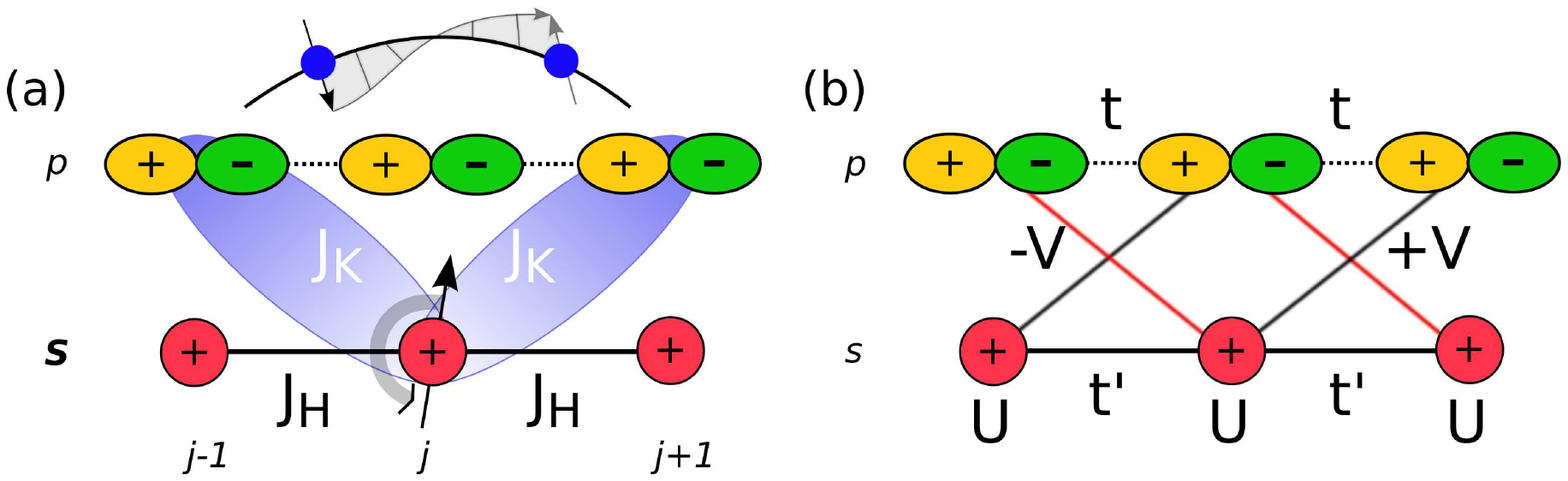}
\protect\caption{(a) Representation of the $p-$wave Kondo-Heisenberg model ($p$-KHM)
in one dimension. The upper leg represents the conduction electron
$p$-band, and the lower leg corresponds to a spin-1/2 Heisenberg
chain. The Kondo exchange $J_{K}$ couples a spin $\mathbf{S}_{j}$
with the $p$-wave spin density in the conduction band {[}see Eq.
(\ref{eq:H_K}){]}. (b) Microscopic model  effectively realizing the
$p$-KHM at low energies, and allowing an experimental implementation in $p-$band optical lattices. At half-filling, the fermionic
sites in the lower-leg behave as spins due to a strong on-site Hubbard
repulsion $U$. The direct hopping across a given rung vanishes due to the different parities of the orbitals. The unusual $p-$wave Kondo interaction in \ref{fig:system}(a) originates in second-order hopping processes in $V$, i.e.,  $J_K = 8V^2/U$ (see Appendix \ref{sec:appendix}). \label{fig:system}}
\end{figure}
In order to gain further intuition into the effect of strong interactions,
recently Alexandrov and Coleman \citep{Alexandrov2014_End_states_in_1DTKI}
proposed an analytically tractable model for a one-dimensional (1D)
TKI, i.e., the ``$p$-wave'' Kondo-Heisenberg model ($p$-KHM), consisting of a chain of spin-1/2 magnetic impurities interacting
with a half-filled one-dimensional electron gas through a Kondo exchange
{[}see Fig. \ref{fig:system}(a){]}. The peculiarity of this model,
which makes it crucially different from other one-dimensional Kondo
lattice models studied previously \citep{zachar_kondo_chain_toulouse,Sikkema97_Spin_gap_in_a_doped_Kondo_chain,zachar_exotic_kondo,Tsunetsugu92_half_filled_Kondo_lattice_1D,Shibata96_Spin_and_charge_gaps_in_the_1DKLM,shibata_kondo_1d,tsunetsugu_kondo_1d,shibata_dmrg_kondo_1d,Zachar01_Staggered_phases_1D_Kondo_Heisenberg_model,Berg10_Pair_density_wave_in_Kondo_Heisenberg_Model,Dobry13_SC_phases_in_the_KH_model,Cho14_Topological_PDW_superconducting_state_in_1D}, is that the Kondo exchange couples to the ``$p$-wave'' conduction-electron
spin density, allowing for effective next-nearest neighbor hopping
processes in the conduction band accompanied by a spin-flip [see Fig. \ref{fig:system}(a)]. %
Using a standard mean-field description \citep{read83,coleman87,newns87},
the above authors found a topologically non-trivial insulating groundstate
(i.e., a class-D insulator \citep{Altland97_Symmetry_classes,Kitaev_TI_classification,Ryu10_Topological_classification})
which hosts magnetic states at the open ends of the chain. Soon after,
two of us studied this system using the Abelian bosonization approach
combined with a perturbative renormalization group analysis, revealing
an unexpected connection to the Haldane phase at low temperatures
\citep{Lobos15_1DTKI}. The Haldane phase is a paradigmatic example
of a strongly interacting topological system, with unique features
such as topologically protected spin-1/2 end-states, non-vanishing
string order parameter, and the breaking of a discrete $Z{}_{2}\times Z_{2}$
hidden symmetry in the groundstate \citep{affleck_klt_short,affleck_klt_long,kennedy_z2z2_haldane}.
The striking results in Ref. \citep{Lobos15_1DTKI} indicate that
1D TKI systems might be much more complex and richer than expected
with the na\"ive mean-field approach, and suggest that they must
be reconsidered from the more general perspective of interacting symmetry-protected
topological phases \citep{Pollmann12_SPT_phases_in_1D,Wang14_Interacting_TIs_in_3D}.

In this Letter we study the groundstate properties of the finite-length
$p$-KHM in one dimension
using the Density Matrix Renormalization Group (DMRG) \citep{White1992,White1993}.
Our results indicate that the system is a Haldane insulator with protected
spin-1/2 end-states and finite string order parameter, therefore
supporting the predictions of Ref. \citep{Lobos15_1DTKI}. %
We also propose that this exotic model could be realized in $p-$band
optical lattices loaded with ultracold Fermi gases, which would allow
for controlled experimental studies of TKIs in the lab.

\textit{Model.} The Hamiltonian of the $p$-KHM is $H=H_{1}+H_{2}+H_{K}$
\citep{Alexandrov2014_End_states_in_1DTKI,Lobos15_1DTKI}, where the
conduction band is represented by a $L$-site tight-binding chain
$H_{1}=-t\sum_{j=1,\sigma}^{L-1}\left(p_{j,\sigma}^{\dagger}p_{j+1,\sigma}+\mbox{H.c.}\right)$
with $p_{j,\sigma}^{\dagger}$ the creation operator of an electron
with spin $\sigma$ at site $j$ with spatial $p-$symmetry {[}upper
leg in Fig. \ref{fig:system}(a){]}. The Hamiltonian $H_{2}=J_{H}\sum_{j=1}^{L-1}\mathbf{S}_{j}\cdot\mathbf{S}_{j+1}$
{[}bottom leg in Fig. \ref{fig:system}(a){]} corresponds to a spin
1/2 Heisenberg chain, and $H_{K}$ is the Kondo exchange coupling
between $H_{1}$ and $H_{2}$ \citep{Alexandrov2014_End_states_in_1DTKI,Lobos15_1DTKI}
\begin{align}
H_{K} & =J_{K}\sum_{j=1}^{L}\mathbf{S}_{j}\ .\ \boldsymbol{\pi}_{j},\label{eq:H_K}
\end{align}
with $J_{K}>0$. This Kondo interaction is unusual in that it couples
the spin $\mathbf{S}_{j}$ in the Heisenberg chain to the ``\textit{p-}wave''
spin density in the fermionic chain at site $j$, defined
as $\boldsymbol{\pi}_{j}\equiv\sum_{\alpha,\beta}\bigl(\frac{p_{j+1,\alpha}^{\dagger}-p_{j-1,\alpha}^{\dagger}}{\sqrt{2}}\bigr)\bigl(\frac{\boldsymbol{\sigma}_{\alpha\beta}}{2}\bigr)\bigl(\frac{p_{j+1,\beta}-p_{j-1,\beta}}{\sqrt{2}}\bigr)$, where $p_{0,\sigma}=p_{L+1,\sigma}=0$ is implied, and where $\boldsymbol{\sigma}_{\alpha\beta}$ is the vector of Pauli matrices.
Eq. (\ref{eq:H_K}) can be written as $H_{K}=H_{K}^{\left(1\right)}+H_{K}^{\left(2\right)}$,
where $H_{K}^{\left(1\right)}=\frac{J_{K}}{2}\sum_{j}\mathbf{S}_{j}\ .\ \left(\mathbf{s}_{j-1}+\mathbf{s}_{j+1}\right)$
contains the coupling of a localized spin with the usual spin-density at site $j\pm 1$ in the conduction band
(where $\mathbf{s}_{j}=\sum_{\alpha,\beta}p_{j,\alpha}^{\dagger}\bigl(\frac{\boldsymbol{\sigma}_{\alpha\beta}}{2}\bigr)p_{j,\beta}$),
and $H_{K}^{\left(2\right)}=-\frac{J_{K}}{2}\sum_{j}\mathbf{S}_{j}\ .\ \left[\sum_{\alpha,\beta}p_{j+1,\alpha}^{\dagger}\bigl(\frac{\boldsymbol{\sigma}_{\alpha\beta}}{2}\bigr)p_{j-1,\beta}+\text{H.c.}\right]$
describes a different type of processes characterized by a non-local
(i.e., next-nearest neighbor) hopping accompanied by a spin-flip. 

We have studied the groundstate properties of $H$ by means of DMRG.
In our implementation we have kept $m=800$ states, which allowed
to achieve truncation errors in the density matrix of the order of $10^{-12}$.
The DMRG method has been used previously to describe the standard
1D Kondo lattice model at half-filling \citep{Tsunetsugu92_half_filled_Kondo_lattice_1D,Shibata96_Spin_and_charge_gaps_in_the_1DKLM,shibata_kondo_1d,tsunetsugu_kondo_1d,shibata_dmrg_kondo_1d},
where a topologically trivial, fully gapped groundstate was obtained.
For the $p$-KHM, where a topological insulator groundstate was predicted
\citep{Alexandrov2014_End_states_in_1DTKI,Lobos15_1DTKI}, there are
no DMRG studies to the best of our knowledge. Intuitively, we expect
that the charge and spin gaps in this model vanish in the limit $J_{K}\rightarrow0$,
as the Hamiltonians $H_{1}$ and $H_{2}$ are separately gapless in
the thermodynamic limit. According to the bosonization analysis in
the limit of small $J_{K}$, both gaps are favored
when the velocities of the gapless spinon excitations described by
$H_{1}$ and $H_{2}$ coincide \citep{Lobos15_1DTKI}. Intuitively,
the term $H_{K}$ becomes more effective to couple the spin degrees
of freedom in $H_{1}$ and $H_{2}$ when they fluctuate coherently
(i.e., same spinon velocities). The spinon velocity in the conduction
band is equal to the tight-binding Fermi velocity $v_{1}=v_{F}=2t$,
and in the Heisenberg chain is $v_{2}=\pi J_{H}/2$ \citep{luther_chaine_xxz,giamarchi_book_1d},
and therefore we conclude that the optimal situation in order to maximize
the effect of $H_{K}$ corresponds to $J_{H}=4t/\pi\approx1.27\ t$,
which we choose in all our subsequent calculations. In what follows,
we characterize the groundstate by analyzing the
charge and spin gaps, the string order parameter, and spin profile along the chain.

\begin{figure}
\includegraphics[bb=10bp 40bp 595bp 450bp,clip,scale=0.42]{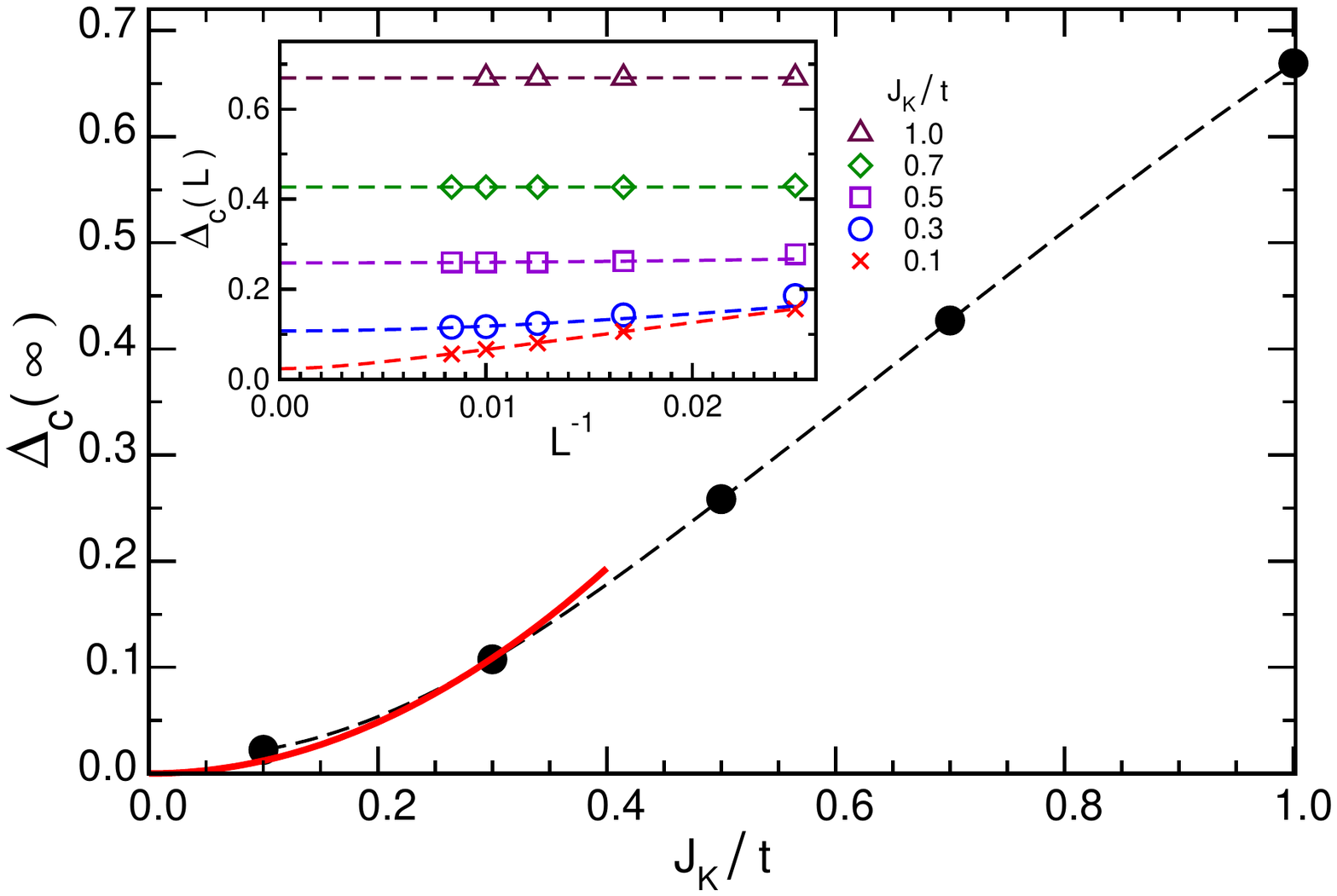}\protect\caption{\label{fig:charge_gap}Charge gap $\Delta_{\text{c}}$ vs $J_{K}$.  
The circles are the values of the gap in the thermodynamical limit $\Delta_{\text{c}}\left(\infty\right)$, obtained after 
finite-size scaling (see inset). The solid (red) line is a fit 
$\Delta_{\text{c}}\left(\infty\right)= \alpha_\text{c} J_{K}^{2}$, valid for small $J_K$, based on the bosonization analysis of Ref. 
\citep{Lobos15_1DTKI}. The dashed lines are a guide to the eye. Inset: Finite-size scaling using the scaling laws $\Delta_{\text{c}}\left(L\right)\approx\Delta_{\text{c}}\left(\infty\right)+\beta_{\text{c}}L^{-2}$
for $J_{K}/t>0.3$ \citep{Shibata96_Spin_and_charge_gaps_in_the_1DKLM,shibata_dmrg_kondo_1d},
and $\Delta_{\text{c}}\left(L\right)\approx\sqrt{\Delta^2_{\text{c}}\left(\infty\right)+\beta_{\text{c}}L^{-2}}$ for $J_{K}/t<0.3$.}
\end{figure}

\textit{Charge gap.} Spin-flip scattering generated upon increasing
$J_{K}$ induces gapped charge- and spin-excitations in the system
at half filling \citep{Alexandrov2014_End_states_in_1DTKI,Lobos15_1DTKI}.
Although these gaps are not direct evidence of the topological nature
of the groundstate, their study is important to characterize the $p-$KHM
insulating phase and to test the predictions of bosonization
\citep{Lobos15_1DTKI}. %
Using the hidden $SU\left(2\right)$ charge pseudo-spin symmetry of
the model at half-filling, we can compute the charge gap of a $L-$supersite system  as $\Delta_{\text{c}}\left(L\right)=E_{0}^{M^{z}=0}\left(N=L+2\right)-E_{0}^{M^{z}=0}\left(N=L\right)$ \citep{shibata_kondo_1d,shibata_dmrg_kondo_1d}.
Here, a ``supersite'' $j$ refers to the combination of a spin $\mathbf{S}_{j}$
and the fermionic site  in each rung.  $M^z$ is the $z$-projection of the total spin in the system, computed as  $M^z=\sum_{j=1}^L \langle T^z_j\rangle$, where $\mathbf{T}_{j}\equiv\mathbf{S}_{j}+\mathbf{s}_{j}$.
Finally,  $E_{0}^{M^{z}}\left(N\right)$ is the groundstate energy of
a system with $N$ electrons in the conduction band, and projection $M^z$.
%
While previous results on the standard 1D Kondo lattice predicted
a linear dependence $\Delta_{\text{c}}\propto J_{K}$ \citep{zachar_kondo_chain_toulouse,shibata_dmrg_kondo_1d},
here the presence of the Heisenberg coupling $J_{H}$ changes this behavior as it cancels the first order contribution
\citep{Lobos15_1DTKI}. %
The leading second-order contribution to $\Delta_{\text{c}}$ can
be physically understood integrating out the ``fast'' spin fluctuations
in the Heisenberg chain, which generate an effective repulsive four-fermion
interaction $U^{\prime}\propto J_{K}^{2}\left\langle \mathbf{S}_{i}.\mathbf{S}_{j}\right\rangle _{H_{2}}$
in the conduction $p$-band. This effective interaction produce umklapp processes which open a Mott insulating gap at half-filling
\citep{giamarchi_book_1d}. In Fig. \ref{fig:charge_gap} we show
the charge gap as a function of $J_{K}$. The system presents important finite-size effects in the limit of small $J_{K}$, and we therefore analyze our results with the scaling law $\Delta_{\text{c}}\left(L\right)\approx\Delta_{\text{c}}\left(\infty\right)+\beta_{\text{c}}L^{-2}$
in the case of large $J_{K}$ ($J_{K}/t>0.3$) \citep{Shibata96_Spin_and_charge_gaps_in_the_1DKLM,shibata_dmrg_kondo_1d},
whereas in the regime of smaller $J_{K}/t<0.3$ the fits improve with the scaling law 
$\Delta_{\text{c}}\left(L\right)\approx\sqrt{\Delta^2_{\text{c}}\left(\infty\right)+\beta_{\text{c}}L^{-2}}$ (see inset in Fig. \ref{fig:charge_gap}).
This fitting procedure allows to extract $\Delta_{\text{c}}\left(\infty\right)$,
the value of the charge gap in the thermodynamic limit, as a function
of $J_{K}$ (see Fig. \ref{fig:charge_gap}). The solid (red) line is a quadratic
law $\Delta_{\text{c}}\left(\infty\right)=\alpha_{\text{c}}J_{K}^{2}$
which fits the data reasonably well at small $J_{K}$, confirming
the dependence predicted by bosonization \citep{Lobos15_1DTKI}. 

\begin{figure}
\includegraphics[bb=10bp 40bp 600bp 450bp,clip,scale=0.42]{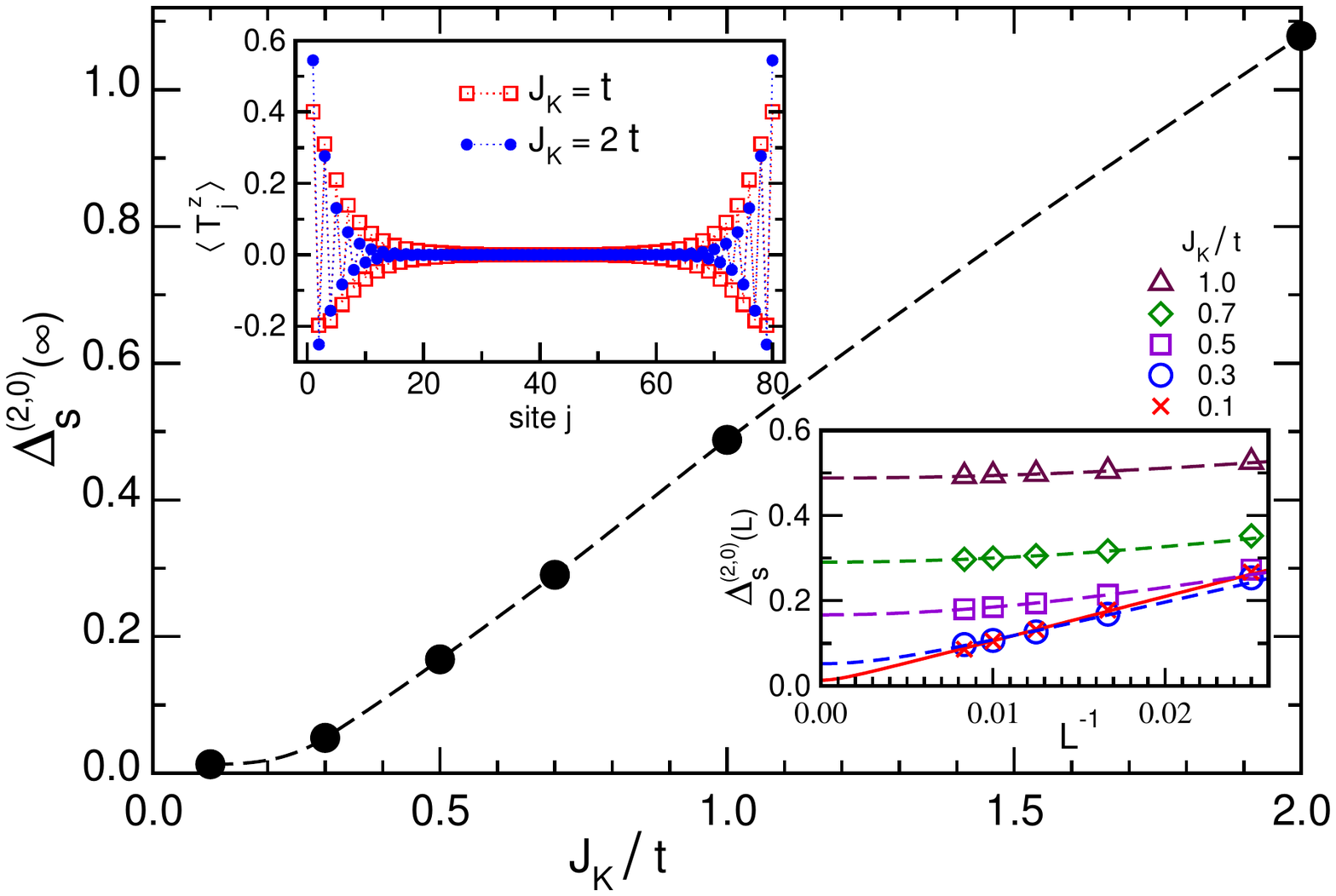}\protect\caption{\label{fig:spin_gap}Spin gap $\Delta_{\text{s}}^{\left(2,0\right)}$ (physically corresponding to the Haldane gap in the system),
as a function of $J_{K}$. The extrapolated
spin gap $\Delta_{\text{s}}^{\left(2,0\right)}\left(\infty\right)$ is obtained after a finite-size scaling analysis (see bottom inset). The dashed lines are a guide to the eye. Top inset: Spatial profile of   $\langle T^z_j \rangle=\langle \psi_{g}^{M^z=1}|T_{j}^{z}|\psi_{g}^{M^{z}=1}\rangle$, i.e., 
the $z$ component of the spin in the supersite $j$, computed with the groundstate of the subspace with total $M^z=1$ for $L=80$. 
The presence of the topologically protected spin-1/2 states at the ends of the chain is clearly seen. Bottom inset: Finite-size scaling using the scaling law $\Delta_{\text{s}}\left(L\right)\approx\sqrt{\Delta^2_{\text{s}}\left(\infty\right)+\beta_{\text{s}}L^{-2}}$.}
\end{figure}

\textit{Spin gap and spin-1/2 end states.} We now focus on the spin
degrees of freedom, where the $p$-KHM has the most interesting properties.
Intuitively, the physics of the problem can be simply understood:
the antiferromagnetic Kondo exchange along the diagonal rungs effectively
forces the spins to align \textit{ferromagnetically} across the rungs,
even in the absence of a direct coupling \citep{Lobos15_1DTKI}. This
situation favors the formation of a local triplet in each supersite,
and the system mimics the properties of the spin-1 Heisenberg chain
\citep{White93_DMRG_S1_Heisenberg_chain} or the ferromagnetic
Kondo lattice model \citep{Tsunetsugu92_half_filled_Kondo_lattice_1D,Garcia04_Spin_order_in_1D_Kondo_and_Hund_lattices},
which are examples of systems realizing an insulating Haldane groundstate.
A hallmark of this phase is the presence of two topologically protected
spin-1/2 magnetic states at the ends of the chain (i.e., $\left|\uparrow\right\rangle _{L}\otimes\left|\uparrow\right\rangle _{R}$,
$\left|\uparrow\right\rangle _{L}\otimes\left|\downarrow\right\rangle _{R}$,
$\left|\downarrow\right\rangle _{L}\otimes\left|\uparrow\right\rangle _{R}$,
and $\left|\downarrow\right\rangle _{L}\otimes\left|\downarrow\right\rangle _{R}$
), which arrange into degenerate triplet and singlet linear combinations.
As a result, the first spin-excitation gap $\Delta_{\text{s}}^{\left(1,0\right)}\left(L\right)\equiv E_{0}^{M^z=1}\left(N=L\right)-E_{0}^{M^z=0}\left(N=L\right)$, 
tends to zero for $L\rightarrow\infty$. For a finite-$L$ chain,
however, the overlap of the end-states wavefunctions removes this
degeneracy exponentially as $\Delta_{\text{s}}^{\left(1,0\right)}\left(L\right)\propto e^{-L/\xi}$,
where $\xi\propto J_{K}^{-1}$ is the localization length for the
magnetic end-states, and the groundstate for $N$ even (odd) corresponds
to the singlet $S=0$ (triplet $S=1$) \citep{White1993}. In our
case, at small $J_{K}$ the localization length $\xi$ becomes of
the order of the system size ($\xi\sim L$), and it was not possible
to obtain a conclusive scaling behavior for $\Delta_{\text{s}}^{\left(1,0\right)}$,
even for the largest systems we have simulated ($L=120$). On the
other hand, the gap $\Delta_{\text{s}}^{\left(2,0\right)}$ (same
definition as above changing $M^z=1 \rightarrow M^z=2$) can be identified with
the Haldane gap of the system, and physically involves spin excitations
which live in the bulk (see Fig. \ref{fig:spin_gap}). In this case,
the scaling analysis is simpler as it is free from edge effects, and
we have used the scaling law 
$\Delta_{\text{s}}\left(L\right)\approx\sqrt{\Delta^2_{\text{s}}\left(\infty\right)+\beta_{\text{s}}L^{-2}}$
for all values of $J_{K}$ (see bottom inset in Fig. \ref{fig:spin_gap}). The values of $\Delta_{\text{s}}\left(\infty\right)$
are shown in Fig. \ref{fig:spin_gap}. In contrast to the case of the charge
gap, here the analytic dependence of $\Delta_{\text{s}}\left(\infty\right)$
on the parameter $J_{K}$ is technically more challenging to obtain
within the bosonization formalism, and is beyond the scope of this work. Nevertheless, our numerical results yield a power-law
dependence $\Delta_{\text{s}}\left(\infty\right)\propto J_{K}^{\nu}$,
with exponent $\nu \cong 2$. 

We next investigate the presence of topologically protected spin-1/2
end-states which, as mentioned before, is a crucial feature of the
open Haldane chain. 
In the upper inset of Fig. \ref{fig:spin_gap} we show a spatial profile
of the $z-$projection of $\mathbf{T}_{j}$, i.e., $\left\langle T_{j}^{z}\right\rangle =\left\langle \psi_{g}^{M^z=1}\right|T_{j}^{z}\left|\psi_{g}^{M^Z=1}\right\rangle $,
where $\left|\psi_{g}^{M^z=1}\right\rangle $ is the groundstate
with total spin $M^z=1$, for $J_{K}/t=1$ and $J_{K}/t=2$. For
these large values of $J_{K}$ (which are beyond the validity of the
bosonization analysis \citep{Lobos15_1DTKI}) the end-states are clearly visible and show a small localization length
$\xi$, a fact that prevents them from overlapping, producing negligible
finite-size effects. Since we are working in the subspace $M^z=1$, and since the spin profile is symmetric under space
inversion, we conclude that the spin accumulation at each end is 1/2,
corresponding to the configuration where the topological spin-1/2
end-states is $\left|\psi_{g}^{M^z=1}\right\rangle \propto \left|\uparrow\right\rangle _{L}\otimes\left|\uparrow\right\rangle _{R}$
\citep{affleck_klt_short,affleck_klt_long}.

\begin{figure}
\includegraphics[bb=25bp 30bp 650bp 420bp,clip,scale=0.45]{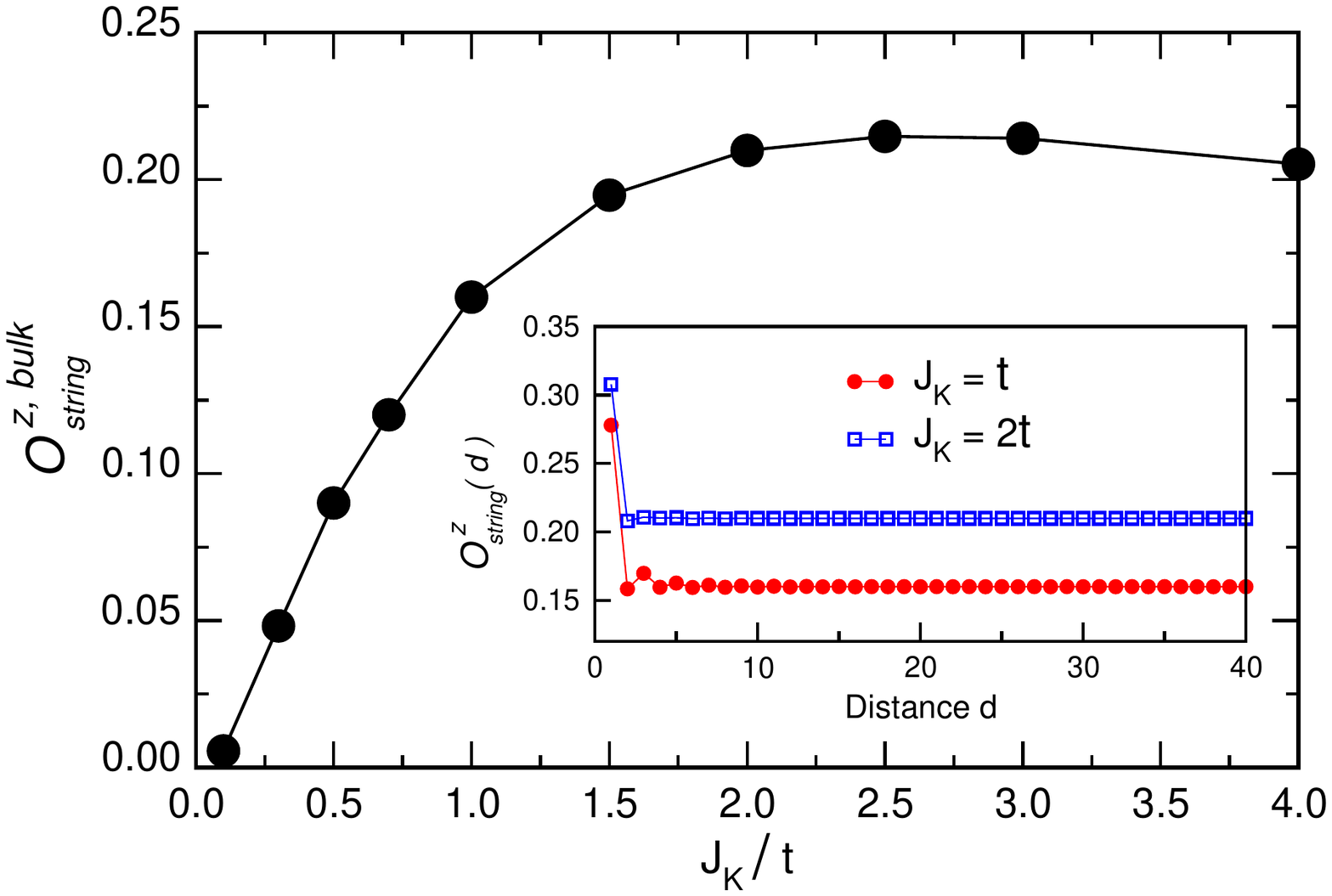}
\protect\caption{String order parameter $\mathcal{O}_{\text{string}}^{z,\text{bulk}}$ vs $J_K$. Throughout the whole studied regime, $\mathcal{O}_{\text{string}}^{z,\text{bulk}}$ remains finite, indicating the
presence of a Haldane-insulating phase. Inset:  Spatial dependence of  $\mathcal{O}_{\text{string}}^{z} \left(d\right)$ vs the distance $d=\left|l-m\right|$, where $l$ and $m$ have been taken symmetrically about the center of the chain. \label{fig:string_order}}
\end{figure}

\textit{String order parameter.} The most fundamental signature of
the Haldane phase is, however, the emergence of a finite string order
parameter \citep{nijs92_roughening}, a quantity deeply connected
to a broken hidden $Z_{2}\times Z_{2}$ symmetry \citep{kennedy_z2z2_haldane}.
This quantity is a smoking-gun for the presence of the Haldane phase,
and therefore is the most important for our present purposes. Using
the above definition of $\mathbf{T}_{j}$,  the string order
parameter is defined as $\mathcal{O}_{\text{string}}^{\alpha}\left(l-m\right)\equiv-\left\langle T_{l}^{\alpha}e^{i\pi\sum_{l<j<m}T_{j}^{\alpha}}T_{m}^{\alpha}\right\rangle $. Due to the $SU\left(2\right)$ spin-symmetry of the model, it is
enough to calculate the computationally simpler component $\alpha=z$.
We have computed $\mathcal{O}_{\text{string}}^{z}\left(l-m\right)$ taking the sites $l$ and $m$ symmetrically about the center of the system in order to minimize the effect of the edges. Note in the inset of Fig. \ref{fig:string_order} that $\mathcal{O}_{\text{string}}^{z} \left(d\right)$ converges rapidly as a function of the distance $d=\left|l-m\right|$ to the 1D-bulk value $\mathcal{O}_{\text{string}}^{z, \text{bulk}}$. In the main Fig. \ref{fig:string_order} we show $\mathcal{O}_{\text{string}}^{z, \text{bulk}}$ vs $J_K$, which remains finite throughout the whole studied regime. This indicates the
presence of a Haldane-insulating phase even beyond the regime of small
$J_{K}$ where the bosonization analysis in Ref. \citep{Lobos15_1DTKI} is valid. 
This result, together with the confirmation of the presence of spin-1/2
end states, are the most important results of this paper, as they
provide conclusive evidence that the $p$-KHM realizes a Haldane insulating
phase. 

\textit{Realization in experimental systems.} Recent experimental
\citep{Wirth11_Superfluidity_in_p_band_optical_lattice,Soltan-Panahi11_QPT_in_multiorbital_optical_lattices,Lewenstein11_Orbital_dance}
and theoretical \citep{Li13_Topological_interacting_ladders,Liu15_Detecting_pi_phase_superfluids}
works on optical lattices with higher orbital $p-$bands suggest that
the $p$-KHM could be realized in the laboratory. More specifically,
the tight-binding Hamiltonian $H_{1}$ could be simulated populating
the first excited energy level with $p-$symmetry at each site in
a 1D optical lattice (see upper leg in Fig. \ref{fig:system}(b)). The Heisenberg chain $H_{2}$ could be simulated 
with a half-filled Hubbard model in the Mott insulating phase \citep{Schneider08_Mott_phase_with_fermons_in_3D_optical_lattice,Joerdens08_Mott_insulator_of_fermionic_atoms_in_an_optical_lattice},
i.e., $H_{\text{Hubbard}}=-t^{\prime}\sum_{\left\langle ij\right\rangle ,\sigma}^{N}\left(s_{i,\sigma}^{\dagger}s_{j,\sigma}+\mathrm{H.c.}\right)+U\sum_{i}^{N}\left(n_{i,\uparrow}^{\left(s\right)}-\frac{1}{2}\right)\left(n_{i,\downarrow}^{\left(s\right)}-\frac{1}{2}\right)$
in the limit $U\gg t^{\prime}$ (see lower leg in Fig. \ref{fig:system}(b)), where $U$ is the on-site Coulomb
interaction which could be tuned via a Feschbach resonance. Here $s_{j,\sigma}^{\dagger}$
is a creation operator of a fermion with spin $\sigma$ at the $s-$orbital
site $j$, and $n_{j,\sigma}^{\left(s\right)}=s_{j,\sigma}^{\dagger}s_{j,\sigma}$
the fermion occupation. %
The unusual Kondo exchange $H_{K}$ %
{} would naturally arise in this situation if a microscopic single-particle
hopping between $H_{1}$ and $H_{\text{Hubbard}}$ is allowed {[}see
Fig. \ref{fig:system}(b){]}. Due to the different parities of the
orbitals involved, the matrix element connecting the sites on the
same rung (i.e., along the vertical direction) vanishes. The leading
contribution therefore corresponds to the matrix element $V$ coupling
$s$ and $p$ orbitals along a \textit{diagonal} rung, i.e., $H_{\text{s-p}}=V\sum_{j,\sigma}s_{j,\sigma}^{\dagger}\left(p_{j+1,\sigma}-p_{j-1,\sigma}\right)+\text{H.c.}$,
where the crucial sign inside the parentheses is a direct consequence
of the $p-$wave nature of the conduction band states. %
The equivalence between the more ``physical'' Hamiltonian $H^{\prime}=H_{\text{Hubbard}}+H_{\text{s-p}}+H_{1}$
and the $p$-KHM can be rigorously shown by the means of a canonical
(i.e., a generalized Schrieffer-Wolff) transformation $\mathcal{T}\equiv e^{i\mathcal{S}}$,
where the operator $\mathcal{S}=\mathcal{S}\left(t^{\prime},V\right)$
must be chosen so as to eliminate first order contributions in $t^{\prime}$
and $V$. The procedure is standard and here we only outline the
main steps (see Appendix \ref{sec:appendix} for details). Assuming
the limit $t^{\prime},V\ll U$, we can expand the exponential in $\mathcal{T}$
and truncate the series at second order in $\left(t^{\prime}/U\right)$
and $\left(V/U\right)$, therefore obtaining $H=\mathcal{T}^{\dagger}H^{\prime}\mathcal{T}\approx H^{\prime}+i\left[\mathcal{S},H^{\prime}\right]+\frac{i^{2}}{2!}\left[\mathcal{S},\left[\mathcal{S},H^{\prime}\right]\right]$.
We choose the transformation to be $\mathcal{S}=-i\left[\left(H_{t^{\prime}}^{+}-H_{t^{\prime}}^{-}\right)+2\left(H_{\text{s-p}}^{+}-H_{\text{s-p}}^{-}\right)\right]/U^{\prime}$,
with $H_{t^{\prime}}^{+}=-t^{\prime}\sum_{\left\langle ij\right\rangle ,\sigma}^{N}n_{i,\bar{\sigma}}^{\left(s\right)}s_{i,\sigma}^{\dagger}s_{j,\sigma}\left(1-n_{j,\bar{\sigma}}^{\left(s\right)}\right)$
and $H_{\text{s-p}}^{+}=V\sum_{i,\sigma}\left[n_{i,\bar{\sigma}}^{\left(s\right)}s_{i,\sigma}^{\dagger}\left(p_{i+1,\sigma}-p_{i-1,\sigma}\right)\right.$$\left.s_{i,\sigma}\left(1-n_{i,\bar{\sigma}}^{\left(s\right)}\right)\right]$,
(with the notation $\left\langle ij\right\rangle $ indicating that
$i$ and $j$ are nearest-neighbor sites), and where $H_{t^{\prime}}^{-}=\left(H_{t^{\prime}}^{+}\right)^{\dagger}$
and $H_{\text{s-p}}^{-}=\left(H_{\text{s-p}}^{+}\right)^{\dagger}$.
It is easy to check that the first order contributions cancel, and
therefore $H_{2}=-\frac{1}{U}\mathcal{P}\left(H_{t}^{-}H_{t}^{+}\right)\mathcal{P}$
and $H_{K}=-\frac{2}{U}\mathcal{P}\left(H_{\text{s-p}}^{-}H_{\text{s-p}}^{+}\right)\mathcal{P}$,
where $\mathcal{P}$ is the projector onto the lowest
subspace of $H_{\text{Hubbard}}$ (see Appendix \ref{sec:appendix}).
The connection between both models is completed identifying the parameters as $J_{H}\equiv4t^{\prime2}/U$ and $J_{K}\equiv8V^{2}/U$.

Note that this proposal is different from other theoretical
proposals to simulate the standard Kondo lattice model in 1D optical
lattices \citep{FossFeig10_Probing_the_Kondo_lattice_model_cold_atoms,FossFeig10_Heavy_fermions_in_optical_lattice}.

\textit{Conclusions. } We have studied the $p$-KHM, a theoretical
``toy-model'' introduced to describe a one-dimensional topological
Kondo insulator, by the means of DMRG, and have calculated various
quantities characterizing the properties of the groundstate at half-filling.
We have shown strong numerical evidence (based on the analysis of
the charge and spin gaps, the spin profile, and the string order
parameter) that the $p$-KHM realizes
a Haldane insulating phase at low temperatures, as predicted in Ref.
\citep{Lobos15_1DTKI}. Our results indicate that the topological
properties of this model fall beyond the scope of the non-interacting
topological classification \citep{Altland97_Symmetry_classes,Kitaev_TI_classification,Ryu10_Topological_classification},
which is unable to reveal the true topological structure of the groundstate.
Finally, we have proposed that the unusual $p-$wave nature of the Kondo
interaction could be physically realized in experiments with ultracold
Fermi gases loaded in $p-$band optical lattices. %

The authors acknowledge support from CONICET-PIP 00389CO.

\bibliographystyle{apsrev_nourl}

\pagebreak{}

\onecolumngrid
\appendix
\section{Derivation of the $p$-KHM by a Canonical Transformation\label{sec:appendix}}

In this Appendix we provide a derivation of the $p$-KHM Hamiltonian $H$
in the main text by the means of a canonical transformation. To that end, we start from the
microscopic Hamiltonian $H^{\prime}$, consisting of a fermionic Hubbard
ladder with $s$ and $p$ orbitals along the legs, and depicted in
Figure 1(b) in the main text:
\begin{align}
H^{\prime}= & H_{\text{Hubbard}}+H_{\text{s-p}}+H_{1},\label{eq:Hprime}\\
H_{\text{Hubbard}}= & -t^{\prime}\sum_{j,\sigma}^{L-1}\left(s_{j,\sigma}^{\dagger}s_{j+1,\sigma}+\mathrm{H.c.}\right)\label{eq:HHubbard}\\
 & +U\sum_{j=1}^{L}\left(n_{j,\uparrow}^{\left(s\right)}-\frac{1}{2}\right)\left(n_{j,\downarrow}^{\left(s\right)}-\frac{1}{2}\right),\\
H_{1}= & -t\sum_{j=1,\sigma}^{L-1}\left(p_{j,\sigma}^{\dagger}p_{j+1,\sigma}+\mbox{H.c.}\right)\label{eq:H1}\\
H_{\text{s-p}}= & V\sum_{j,\sigma}s_{j,\sigma}^{\dagger}\left(p_{j+1,\sigma}-p_{j-1,\sigma}\right)+\text{H.c.},\label{eq:Hsp}
\end{align}
Note that the system has electron-hole symmetry. Here, $s_{j,\sigma}^{\dagger}$
creates a fermion with spin projection $\sigma$
at site $j$ in the Hubbard leg and $n_{j\sigma}^{\left(s\right)}\equiv s_{j,\sigma}^{\dagger}s_{j,\sigma}$
is the corresponding fermion-number operator. The operator $p_{j,\sigma}^{\dagger}$ creates a fermion with spin $\sigma$ at site $j$ in the $p-$orbital conduction band, represented by a simple tight-binding
model $H_{1}$. The term $H_{\text{s-p}}$ couples the two fermionic
legs, and due to the symmetry properties of the $s-$ and $p-$orbitals,
the direct hopping across the rungs is zero. Therefore, the most important
hopping process occurs between a fermion $s_{j,\sigma}$ and the linear
superposition with $p-$wave symmetry $\propto \left(p_{j+1,\sigma}-p_{j-1,\sigma}\right)$
in the conduction band. 

The idea is to derive an effective low-energy model in the limit $U\gg\left\{ t^{\prime},V\right\} $.
To that end, we split the Hamiltonian $H^{\prime}$ into

\begin{align}
H^{\prime} & =H_{t^{\prime}}+H_{\text{s-p}}+H_{U}+H_{1},\label{eq:H2}
\end{align}
where
\begin{align}
H_{t^{\prime}} & =-t^{\prime}\sum_{\left\langle ij\right\rangle ,\sigma}^{L}\left(s_{i,\sigma}^{\dagger}s_{j,\sigma}+\mathrm{H.c.}\right),\label{eq:Ht}\\
H_{U} & =U\sum_{j}^{L}\left(n_{j,\uparrow}-\frac{1}{2}\right)\left(n_{j,\downarrow}-\frac{1}{2}\right).\label{eq:HU}
\end{align}
The first two terms in (\ref{eq:H2}) will be considered as perturbations
to $H_{U}$, in the regime $\left\{ t^{\prime},V\right\} \ll U$. 

We now start from the atomic limit in the Hubbard leg, i.e., $t^{\prime}=V=0$,
and identify the atomic singly-occupied states $\left|\sigma_{j}\right\rangle =s_{j,\sigma}^{\dagger}\left|0\right\rangle $
($\sigma=\uparrow,\downarrow$) as forming the lowest-energy subspace
at site $j$, while the $\left|0_{j}\right\rangle $ (empty) and $\left|d_{j}\right\rangle =s_{j,\uparrow}^{\dagger}s_{j,\downarrow}^{\dagger}\left|0\right\rangle $
(doubly-occupied) form the excited subspace. We now introduce projectors
onto each of the 4 atomic states:

\begin{align}
\mathcal{P}_{j,0} & =\left(1-n_{j,\uparrow}^{\left(s\right)}\right)\left(1-n_{j,\downarrow}^{\left(s\right)}\right),\label{eq:P0}\\
\mathcal{P}_{j,d} & =n_{j,\uparrow}^{\left(s\right)}n_{j,\downarrow}^{\left(s\right)},\label{eq:Pd}\\
\mathcal{P}_{j,\uparrow} & =n_{j,\uparrow}^{\left(s\right)}\left(1-n_{j,\downarrow}^{\left(s\right)}\right),\label{eq:Pup}\\
\mathcal{P}_{j,\downarrow} & =n_{j,\downarrow}^{\left(s\right)}\left(1-n_{j,\uparrow}^{\left(s\right)}\right).\label{eq:Pdown}
\end{align}
Note that while all projectors commute with $H_{U}$, the kinetic
terms $H_{t^{\prime}}$ and $H_{\text{s-p}}$ cause transitions among
subspaces. Using that $\mathbf{1}_{j}=\sum_{\alpha}\mathcal{P}_{j,\alpha}$,
we can write the kinetic terms as $H_{t^{\prime}}=\left(\sum_{i,\alpha}\mathcal{P}_{i,\alpha}\right)H_{t^{\prime}}\left(\sum_{j,\beta}\mathcal{P}_{j,\beta}\right)=H_{t^{\prime}}^{+}+H_{t^{\prime}}^{-}+H_{t^{\prime}}^{0},$
and $H_{\text{s-p}}=\left(\sum_{i,\alpha}\mathcal{P}_{i,\alpha}\right)H_{\text{s-p}}\left(\sum_{j,\beta}\mathcal{P}_{j,\beta}\right)=H_{\text{s-p}}^{+}+H_{\text{s-p}}^{-}+H_{\text{s-p}}^{0},$
where%

\begin{align}
H_{t^{\prime}}^{+} & =-t^{\prime}\sum_{\left\langle ij\right\rangle ,\sigma}^{N}\left[n_{i,\bar{\sigma}}^{\left(s\right)}s_{i,\sigma}^{\dagger}s_{j,\sigma}\left(1-n_{j,\bar{\sigma}}^{\left(s\right)}\right)+n_{j,\bar{\sigma}}^{\left(s\right)}s_{j,\sigma}^{\dagger}s_{i,\sigma}\left(1-n_{i,\bar{\sigma}}^{\left(s\right)}\right)\right],\label{eq:Htplus}\\
H_{t^{\prime}}^{-} & =-t^{\prime}\sum_{\left\langle ij\right\rangle ,\sigma}^{N}\left[\left(1-n_{j,\bar{\sigma}}^{\left(s\right)}\right)s_{j,\sigma}^{\dagger}s_{i,\sigma}n_{i,\bar{\sigma}}^{\left(s\right)}+\left(1-n_{i,\bar{\sigma}}^{\left(s\right)}\right)s_{i,\sigma}^{\dagger}s_{j,\sigma}n_{j,\bar{\sigma}}^{\left(s\right)}\right],\label{eq:Htminus}\\
H_{\text{s-p}}^{+} & =V\sum_{i,\sigma}\left[n_{i,\bar{\sigma}}^{\left(s\right)}s_{i,\sigma}^{\dagger}\left(p_{i+1,\sigma}-p_{i-1,\sigma}\right)+\left(p_{i+1,\sigma}^{\dagger}-p_{i-1,\sigma}^{\dagger}\right)s_{i,\sigma}\left(1-n_{i,\bar{\sigma}}^{\left(s\right)}\right)\right],\label{eq:Hmixplus}\\
H_{\text{s-p}}^{-} & =V\sum_{i,\sigma}\left[\left(p_{i+1,\sigma}^{\dagger}-p_{i-1,\sigma}^{\dagger}\right)s_{i,\sigma}n_{i,\bar{\sigma}}^{\left(s\right)}+\left(1-n_{i,\bar{\sigma}}^{\left(s\right)}\right)s_{i,\sigma}^{\dagger}\left(p_{i+1,\sigma}-p_{i-1,\sigma}\right)\right].\label{eq:Hmixminus}
\end{align}
Physically, the term with supraindex ``$+$'' produce
transitions from the lowest subspace to the excited subspace, while
those with ``$-$'' restore excited states to the lowest subspace. On
the other hand, the terms labelled with ``$0$'' do not change the subspace,
and since we assume a half-filled conduction band, they will identically vanish and
it is not necessary to write them explicitly here. We now note the
following important relations

\begin{align}
H_{t^{\prime}}^{-} & =\left(H_{t^{\prime}}^{+}\right)^{\dagger},\label{eq:hermitian_relation}\\
H_{\text{s-p}}^{-} & =\left(H_{\text{s-p}}^{+}\right)^{\dagger},
\end{align}
which will be useful in what follows.

We now introduce a canonical transformation in Eq. (\ref{eq:Hprime}),
such that in the transformed representation we simultaneously get
rid of the terms at first order in $t^{\prime}$ and $V$:
\begin{align}
H & =e^{i\mathcal{S}}H^{\prime}e^{-i\mathcal{S}}.\label{eq:Heff_canonical_transf}\\
 & =H^{\prime}+i\left[\mathcal{S},H^{\prime}\right]+\frac{i^{2}}{2!}\left[\mathcal{S},\left[\mathcal{S},H^{\prime}\right]\right]+\dots\label{eq:expansion}
\end{align}
We want to choose $\mathcal{S}$ in such a way that $H$ does not
connect different Hubbard subbands. Note that this cannot be achieved
at infinite order in the expansion in powers of $\mathcal{S}$ in
Eq. (\ref{eq:expansion}), but we will be content if we can eliminate
the contributions at order $\mathcal{O}\left(t^{\prime}\right)$ and
$\mathcal{O}\left(V\right)$ that mix the subbands. We now write the
expansion in Eq. (\ref{eq:expansion}) in the more suggestive form

\begin{align}
H & =H_{t^{\prime}}^{+}+H_{t^{\prime}}^{-}+H_{\text{s-p}}^{+}+H_{\text{s-p}}^{-}+i\left[\mathcal{S},H_{U}\right]\label{eq:expansion_a}\\
 & +H_{1}+H_{U}+i\left[\mathcal{S},H_{t^{\prime}}^{+}+H_{t^{\prime}}^{-}\right]+i\left[\mathcal{S},H_{\text{s-p}}^{+}+H_{\text{s-p}}^{-}\right]\label{eq:expansion_b}\\
 & +\frac{i^{2}}{2!}\left[\mathcal{S},\left[\mathcal{S},H_{U}\right]\right]+\text{(other less important terms})\label{eq:expansion_c}
\end{align}
We will require that the first line (\ref{eq:expansion_a}) in the
above equation vanishes. It is then clear that $\mathcal{S}$ must
be $\mathcal{O}\left(t^{\prime}/U\right)\sim\mathcal{O}\left(V/U\right)$.
Using the following results 

\begin{align}
\left[n_{i,\bar{\sigma}}^{\left(s\right)}s_{i,\sigma}^{\dagger}s_{j,\sigma}\left(1-n_{j,\bar{\sigma}}^{\left(s\right)}\right),\left(n_{i,\sigma}^{\left(s\right)}-\frac{1}{2}\right)\left(n_{i,\bar{\sigma}}^{\left(s\right)}-\frac{1}{2}\right)\right] & =\left[n_{i,\bar{\sigma}}^{\left(s\right)}s_{i,\sigma}^{\dagger}s_{j,\sigma}\left(1-n_{j,\bar{\sigma}}^{\left(s\right)}\right),\left(n_{j,\sigma}^{\left(s\right)}-\frac{1}{2}\right)\left(n_{j,\bar{\sigma}}^{\left(s\right)}-\frac{1}{2}\right)\right]\label{eq:relation1}\\
 & =-\frac{1}{2}n_{i,\bar{\sigma}}^{\left(s\right)}s_{i,\sigma}^{\dagger}s_{j,\sigma}\left(1-n_{j,\bar{\sigma}}^{\left(s\right)}\right),\label{eq:relation2}\\
\left[n_{i,\bar{\sigma}}^{\left(s\right)}s_{i,\sigma}^{\dagger}\left(p_{i+1,\sigma}-p_{i-1,\sigma}\right),\left(n_{i,\sigma}^{\left(s\right)}-\frac{1}{2}\right)\left(n_{i,\bar{\sigma}}^{\left(s\right)}-\frac{1}{2}\right)\right] & =-\frac{1}{2}n_{i,\bar{\sigma}}^{\left(s\right)}s_{i,\sigma}^{\dagger}\left(p_{i+1,\sigma}-p_{i-1,\sigma}\right),\label{eq:relation3}\\
\left[\left(p_{i+1,\sigma}^{\dagger}-p_{i-1,\sigma}^{\dagger}\right)s_{i,\sigma}\left(1-n_{i,\bar{\sigma}}^{\left(s\right)}\right),\left(n_{i,\sigma}^{\left(s\right)}-\frac{1}{2}\right)\left(n_{i,\bar{\sigma}}^{\left(s\right)}-\frac{1}{2}\right)\right] & =-\frac{1}{2}\left(p_{i+1,\sigma}^{\dagger}-p_{i-1,\sigma}^{\dagger}\right)s_{i,\sigma}\left(1-n_{i,\bar{\sigma}}^{\left(s\right)}\right),\label{eq:relation4}
\end{align}
it is easy to check that

\begin{align}
\left[H_{t^{\prime}}^{\pm},H_{U}\right] & =\mp UH_{t^{\prime}}^{\pm},\label{eq:commutator_HtHU}\\
\left[H_{\text{s-p}}^{\pm},H_{U}\right] & =\mp\frac{U}{2}H_{\text{s-p}}^{\pm}.\label{eq:commutator_HmixHU}
\end{align}
Then, it follows that the choice

\begin{align}
\mathcal{S} & =-\frac{i}{U}\left(H_{t^{\prime}}^{+}-H_{t^{\prime}}^{-}\right)-\frac{2i}{U}\left(H_{\text{s-p}}^{+}-H_{\text{s-p}}^{-}\right),\label{eq:S_definition}
\end{align}
exactly cancels line (\ref{eq:expansion_a}). 

The relevant part of the Hamiltonian at low energies is then obtained
projecting $H$ onto the lowest Hubbard subband. This is formally
done applying the projector $\mathcal{P}=\sum_{i}\left(\mathcal{P}_{i,\uparrow}+\mathcal{P}_{i,\downarrow}\right)$,
which eliminates certain terms in Eqs. (\ref{eq:expansion_b}) and
(\ref{eq:expansion_c}). The resulting effective Hamiltonian at lowest
order in $t/U$ and $V/U$ is therefore%

\begin{align}
H & =H_{1}+\mathcal{P}\left\{ H_{U}+i\left[\mathcal{S},H_{t^{\prime}}^{+}+H_{t^{\prime}}^{-}\right]+i\left[\mathcal{S},H_{\text{s-p}}^{+}+H_{\text{s-p}}^{-}\right]+\frac{i^{2}}{2!}\left[\mathcal{S},\left[\mathcal{S},H_{U}\right]\right]\right\} \mathcal{P},\nonumber \\
 & =H_{1}+H_{U}-\frac{1}{U}\mathcal{P}\left(H_{t^{\prime}}^{-}H_{t^{\prime}}^{+}\right)\mathcal{P}-\frac{2}{U}\mathcal{P}\left(H_{\text{s-p}}^{-}H_{\text{s-p}}^{+}\right)\mathcal{P}.\label{eq:Heff}
\end{align}
We now replace the expressions for $H_{t^{\prime}}^{\pm}$ and $H_{\text{s-p}}^{\pm}$
{[}Eqs. (\ref{eq:Htplus})-(\ref{eq:Hmixminus}){]} into the above
equation and obtain%
{} %
\begin{align}
\mathcal{P}\left(H_{t^{\prime}}^{-}H_{t^{\prime}}^{+}\right)\mathcal{P} & =\left(t^{\prime}\right)^{2}\sum_{i,\sigma}^{N}\left[\mathcal{P}_{i,\bar{\sigma}}\mathcal{P}_{i+1,\sigma}+\mathcal{P}_{i,\sigma}\mathcal{P}_{i+1,\bar{\sigma}}-s_{i+1,\sigma}^{\dagger}s_{i+1,\bar{\sigma}}s_{i,\bar{\sigma}}^{\dagger}s_{i,\sigma}-s_{i+1,\bar{\sigma}}^{\dagger}s_{i+1,\sigma}s_{i,\sigma}^{\dagger}s_{i,\bar{\sigma}}\right]\label{eq:projection1}\\
\mathcal{P}\left(H_{\text{s-p}}^{-}H_{\text{s-p}}^{+}\right)\mathcal{P} & =2V^{2}\sum_{i,\sigma}n_{i,\sigma}^{\left(s\right)}-V^{2}\sum_{i,\sigma}\left[\left(n_{i,\sigma}^{\left(s\right)}-n_{i,\bar{\sigma}}^{\left(s\right)}\right)\left(p_{i+1,\sigma}^{\dagger}-p_{i-1,\sigma}^{\dagger}\right)\left(p_{i+1,\sigma}-p_{i-1,\sigma}\right)\right.\nonumber \\
 & \left.+s_{i,\bar{\sigma}}^{\dagger}s_{i,\sigma}\left(p_{i+1,\sigma}^{\dagger}-p_{i-1,\sigma}^{\dagger}\right)\left(p_{i+1,\bar{\sigma}}-p_{i-1,\bar{\sigma}}\right)+s_{i,\sigma}^{\dagger}s_{i,\bar{\sigma}}\left(p_{i+1,\bar{\sigma}}^{\dagger}-p_{i-1,\bar{\sigma}}^{\dagger}\right)\left(p_{i+1,\sigma}-p_{i-1,\sigma}\right)\right]\label{eq:projection2}
\end{align}
Using that $\mathcal{P}_{i,\uparrow}\mathcal{P}_{i+1,\downarrow}+\mathcal{P}_{i,\downarrow}\mathcal{P}_{i+1,\uparrow}=-2S_{i}^{z}S_{i+1}^{z}+n_{i}^{\left(s\right)}n_{i+1}^{\left(s\right)}/2$,
and the Schwinger-fermion representation 

\begin{align}
S_{i}^{z} & =\frac{n_{i,\uparrow}^{\left(s\right)}-n_{i,\downarrow}^{\left(s\right)}}{2},\label{eq:Sz}\\
S_{i}^{+} & =s_{i,\uparrow}^{\dagger}s_{i,\downarrow},\label{eq:Splus}\\
S_{i}^{-} & =s_{i,\downarrow}^{\dagger}s_{i,\uparrow},\label{eq:Sminus}
\end{align}
is a faithful representation of a spin-1/2 operator, we can write
the effective Hamiltonian as
\begin{align}
H & =H_{U}+H_{1}+\frac{4\left(t^{\prime}\right)^{2}}{U}\sum_{i}^{N}\left[S_{i}^{z}S_{i+1}^{z}+\frac{S_{i+1}^{+}S_{i}^{-}+S_{i+1}^{-}S_{i}^{+}}{2}-\frac{1}{4}\right]\nonumber \\
 & +\frac{8V^{2}}{U}\sum_{i}S_{i}^{z}\frac{\left(p_{i+1,\uparrow}^{\dagger}-p_{i-1,\uparrow}^{\dagger}\right)\left(p_{i+1,\uparrow}-p_{i-1,\uparrow}\right)-\left(p_{i+1,\downarrow}^{\dagger}-p_{i-1,\downarrow}^{\dagger}\right)\left(p_{i+1,\downarrow}-p_{i-1,\downarrow}\right)}{2}\nonumber \\
 & +\frac{8V^{2}}{U}\sum_{i}\left[\frac{S_{i}^{+}\left(p_{i+1,\downarrow}^{\dagger}-p_{i-1,\downarrow}^{\dagger}\right)\left(p_{i+1,\uparrow}-p_{i-1,\uparrow}\right)}{2}+\frac{S_{i}^{-}\left(p_{i+1,\uparrow}^{\dagger}-p_{i-1,\uparrow}^{\dagger}\right)\left(p_{i+1,\downarrow}-p_{i-1,\downarrow}\right)}{2}\right]\label{eq:Heff_final}
\end{align}
where we have neglected the constant $\frac{2V^{2}}{U}\sum_{i,\sigma}n_{i,\sigma}^{\left(s\right)}$.
Defining the effective parameters

\begin{align}
J_{H} & \equiv\frac{4\left(t^{\prime}\right)^{2}}{U},\label{eq:JH}\\
J_{K} & \equiv\frac{8V^{2}}{U},\label{eq:JK}
\end{align}
we note that this Hamiltonian corresponds to the $p$-KHM
considered by Alexandrov and Coleman \citep{Alexandrov2014_End_states_in_1DTKI}.
\end{document}